\documentclass[12pt]{iopart}
\usepackage[utf8]{inputenc}
\usepackage[T1]{fontenc} 
\usepackage{ebgaramond}
\usepackage{graphicx} 
\usepackage{bm} 
\usepackage{times}
\usepackage{mathptmx}
\usepackage{xcolor}
\usepackage{soul}
\usepackage{subcaption}
\usepackage{hyperref}
\usepackage{multirow}

\expandafter\let\csname equation*\endcsname\relax
\expandafter\let\csname endequation*\endcsname\relax
\usepackage{amsmath}
\usepackage{amssymb}
\usepackage{amsfonts}
\usepackage{cleveref} 

\begin{document}

\title[\JPCM]{DMRG and Monte Carlo studies of $\textrm{CrI}_3$ magnetic phases and the phase transition}
\author{Bartosz Rzepkowski$^1$, Michał Kupczyński$^1$, Paweł Potasz$^2$, Arakdiusz Wójs$^1$}
\address{$^1$ Department of Theoretical Physics, 
Faculty of Fundamental Problems of Technology,
Wrocław University of Science and Technology,
PL-50370 Wrocław, Poland}
\address{$^2$Institute of Physics, Faculty of Physics, Astronomy and Informatics, Nicolaus Copernicus University, Grudziadzka 5, 87-100 Toru\'n, Poland}

\ead{bartosz.rzepkowski@pwr.edu.pl}

\begin{abstract}
    The monolayer of $\textrm{CrI}_3$ has been reported to exhibit the ferromagnetic order, with a Curie temperature of $45 K$ and off-plane easy axis, which has attracted much attention in the community of condensed matter physics. Using the Density Matrix Renormalization Group method, we investigate the role of correlation effects and using classical Monte Carlo simulations analyze the nature of phase transitions in the XXZ Hamiltonian on a honeycomb lattice, which can effectively model $\textrm{CrI}_3$. We show, that the magnetic ordering of the Hamiltonian's ground state can be well approximated by classical models in a wide range of the anisotropy parameter space. Using classical Monte Carlo simulations we estimate the Curie temperature of $49.7 K$ which is in good agreement with experimental result.
\end{abstract}

\submitto{\JPCM}
\maketitle

\section{Introduction}
In recent years, the wide class of two-dimensional van der Waals crystals, which includes semiconductors, superconductors, semi-metals, topological insulators, charge density waves materials, ferroelectrics and magnetics, has been extensively studied both by theoretical and experimental methods \cite{Geim.Grigorieva.2013,Kin.Lee.2010,Wang.Kalantar.2012, Chhowalla.Shin.2013,Lu.Zheliuk.2015,Xi.Zhao.2016,Ugeda.Bradley.2016,Novoselov.Mishchenko.2016,Ren.Qiao.2016,Bieniek.Wozniak.2017,Wang.Chernikov.2018,Brzezinska.Bieniek.2018,Masrour.Hlil.2013,Masrour.Hlil.2014,Masrour.Hlil1.2014,Masrour.Hlil1.2013}. 
These materials have many potential applications in novel approaches towards electronics, like spintronics, valleytronics, and optoelectronics \cite{Wang.Kalantar.2012, Splendiani.Sun.2010, Mak.Shan.2016,Schaibley.Yu.2016,Mak.Xiao.2018, Mennel.Symonowicz.2020}.
Moreover, in such materials the long range magnetic order has been observed \cite{Zhonge.Seyler.2017,Samarth.2017,Sander.Valenzuela.2017, Cardoso.Soriano.2018, Shabbir.Nadeem.2018,Miao.Xu.2018,Gibertini.Koperski.2018, Jiang.Wang.2018, Burch.Mandrus.2018}. 
The ferromagnetic order has been found in monolayers of $\textrm{CrI}_3$ and $\textrm{Cr}_2 \textrm{Ge}_2 \textrm{Te}_6$ \cite{Gong.Li.2017,Huang.Clark.2017}, and anti-ferromagnetic order in the monolayer $\textrm{Fe} \textrm{PS}_3$ \cite{Wang.Du.2016, Lee.Lee.2016}.
Also, topologically protected magnetic skyrmions were predicted to appear in these materials \cite{Behera.Chowdhury.2019}.
The magnetic order has also been foreseen and later confirmed experimentally in many  other materials \cite{Ma.Dai.2012, Sachs.Wehling.2013,Sivadas.Daniels.2015,Chittari.Park.2016,Gong.Li.2017, Bonilla.Kolekar.2017,Tan.Lee.2018, Lin.Choi.2018, Otrokov.Rusinov.2019,Tang.Sun.2019, Chittari.Lee.2020,Kargar.Coleman.2020,Yang.Fan.2020,Subhan.Hong.2020}.

It has been shown by Mermin and Wagner that in two-dimensional systems with spin-rotational symmetry and short-range exchange interactions, the long-range magnetic order cannot exist \cite{Mermin.Wagner.1966}. 
Thus, the description of magnetic properties of 2D crystals is focused on the mechanism of breaking spin-rotational invariance. In general, it can be achieved by dipolar interactions, single ion anisotropy and anisotropy of the exchange interactions.

The magnetic order in monolayer of $\textrm{CrI}_3$ can be described by the spin $S = 3/2$ XXZ model with a single ion anisotropy on the honeycomb lattice \cite{Lado.Fernandes.2017,Pizzochero.Yadav.2020}. 
This material is  ferromagnet with off-plane spin orientation below the Curie temperature $T_c = 45K $ \cite{Huang.Clark.2017}.
The proposed Hamiltonian exhibits four possible types of classical magnetic order: in-plane ferromagnetic, in-plane anti-ferromagnetic, off-plane ferromagnetic and off-plane anti-ferromagnetic.  
Appropriate parameters for $\textrm{CrI}_3$ have been determined using \textit{ab-initio} methods \cite{Lado.Fernandes.2017,Pizzochero.Yadav.2020}.
Other phases can be achieved in $\textrm{CrI}_3$ by introducing defects \cite{Pizzochero.2020}, strain \cite{Webster.Yan.2018,Zheng.Zhao.2018} and charge doping \cite{Zheng.Zhao.2018,Jiang.Li.2018}, which effectively enhance magnetic anisotropy. 

Although multilayer and bulk $\textrm{CrI}_3$ have already been studied \cite{Wang.2021,McGuire.2015,Tomarchio.2021}, to the best of our knowledge the monolayer system has not been analyzed for magnetic phase stability as a function of anisotropy parameters. In this work we address this gap, because monolayer $\textrm{CrI}_3$ is a building block for the more complex systems mentioned above, and full understanding of its properties might facilitate their analysis in the future. For that purpose we use the Density Matrix Renormalization Group (DMRG) to obtain the groundstate of the spin Hamiltonian at zero temperature in a wide range of the parameter space. By comparison with classical predictions, the effect of quantum correlations is determined. At finite temperature and realistic parameters, we perform Monte Carlo calculations which allows us to estimate the temperature of phase transition.

The paper is organized as follows. In \cref{sec:models} we present the explicit form of quantum model and its classical approximation, which are being studied. The details on how DMRG simulations were performed, along with obtained results, are given in \cref{sec:dmrg}. In \cref{sec:monte_carlo} we provide predictions of the Curie temperature for isolated $\textrm{CrI}_3$ acquired from Monte Carlo calculations. Finally, \cref{sec:conclusions} concludes the paper.

\section{Methodology} \label{sec:models}
\subsection{Effective spin Hamiltonian}

\begin{figure}[h]
    \begin{subfigure}{0.5\textwidth}
        \centering
        \includegraphics[width=0.52\linewidth]{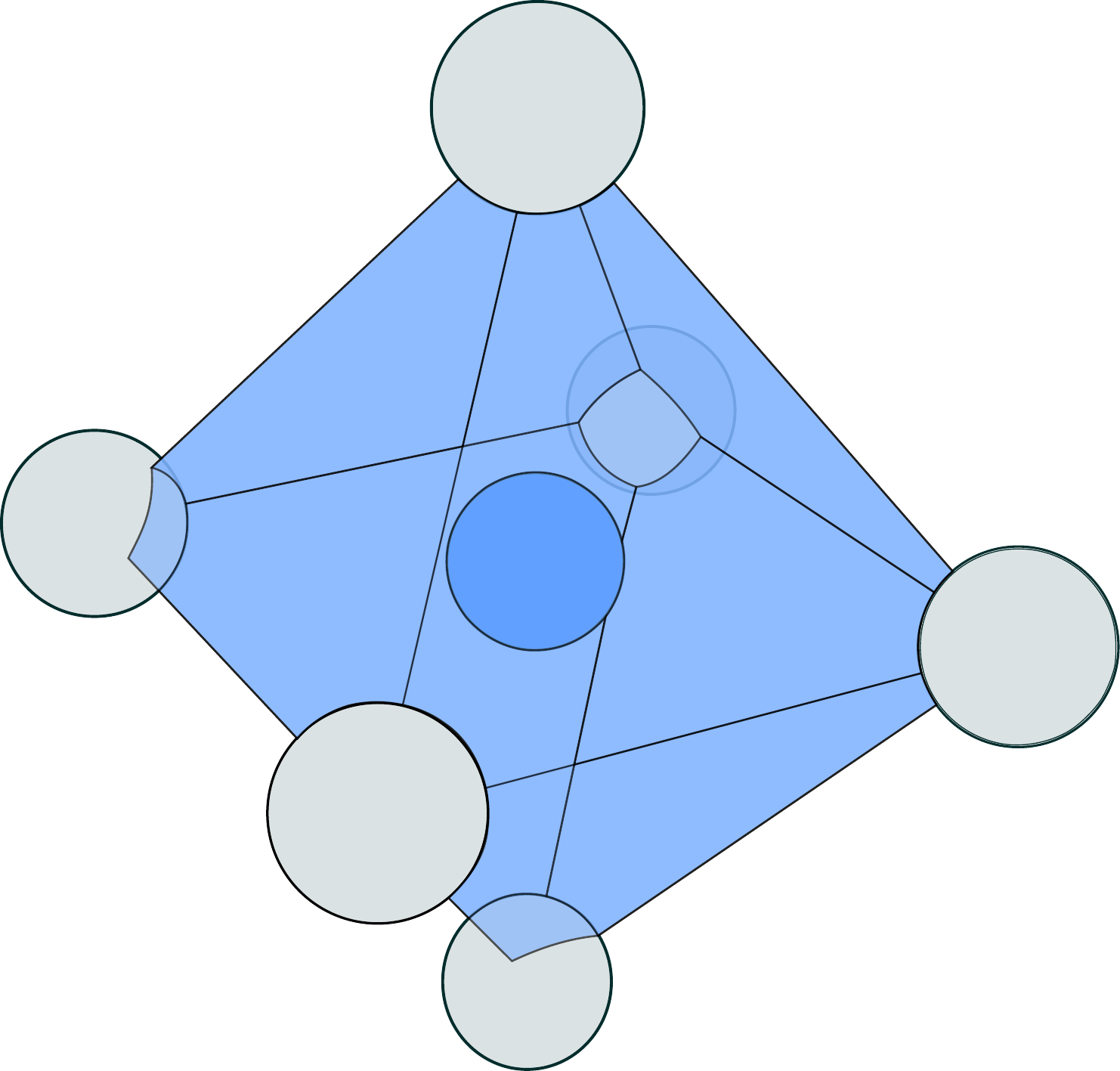} 
        \caption{}
        \label{fig:octahedron}
    \end{subfigure}
    \begin{subfigure}{0.5\textwidth}
        \includegraphics[width=0.93\linewidth]{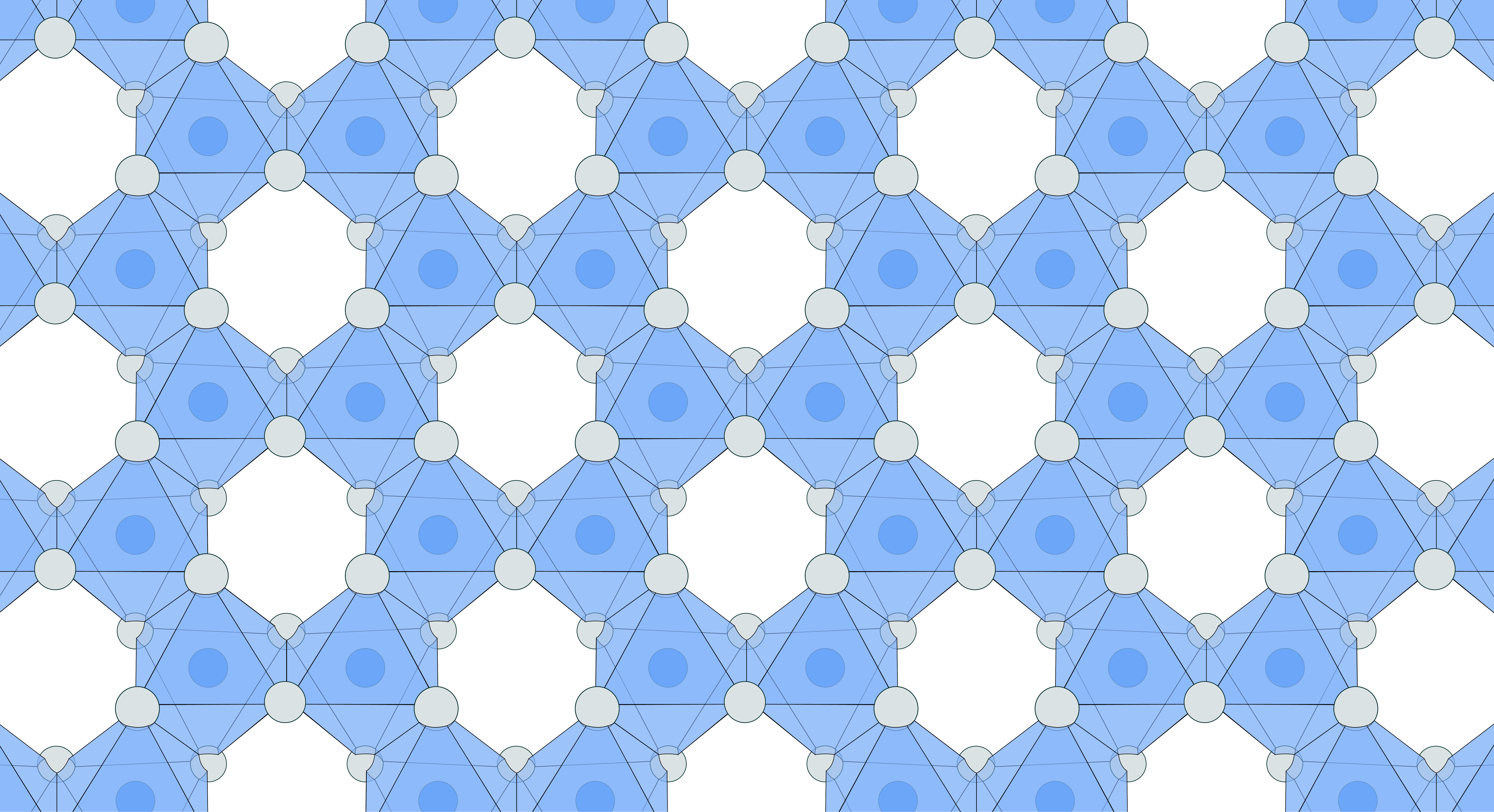}
        \caption{}
        \label{fig:lattice}
    \end{subfigure}

    \caption{(a) Chromium atom with its octahedral iodine environment. (b) The honeycomb crystalline lattice of $\textrm{CrI}_3$.}
\end{figure}

The crystalline structure of $\textrm{CrI}_3$ consists of chromium atoms arranged on a honeycomb lattice. Each chromium atom has an octahedron-shaped environment with iodine atoms lying on its nodes. The chromium atom with its environment is depicted in \cref{fig:octahedron}, while the whole lattice structure of the monolayer $\textrm{CrI}_3$ is shown in \cref{fig:lattice}.

It has been shown, that the XXZ model with the spin $S = 3/2$ on the honeycomb lattice can be used to describe the behavior of two-dimensional $\textrm{CrI}_3$ \cite{Lado.Fernandes.2017}. The effective Hamiltonian can be explicitly written as

\begin{equation} \label{eq:hamiltonian}
    H = - \left (  J \sum_{\left<i,i'\right>, i < i'} \bar{S}_i \cdot \bar{S}_{i'} + \lambda \sum_{\left<i,i'\right>, i < i'} S_i^z S_{i'}^z + \sum_i D(S_i^z )^2 \right),
\end{equation}

where $Z$ axis was chosen as the off-plane direction. The Heisenberg isotropic exchange interaction between spin particles corresponds to the $J$ parameter. The anisotropic symmetric exchange, originating from the spin-orbit interaction of the ligand I atoms across the $\simeq 90$ degree Cr-I-Cr bonds, is represented by $\lambda$. Finally, single ion anisotropy arising from the interaction between the spin–orbit coupling and the distorted octahedral environment of Cr atoms, is described by the $D$ term. 

The minus sign standing at the beginning of \cref{eq:hamiltonian} was chosen to favor the ferromagnetic interactions for $J > 0$, and the alignment of spins in the off-plane axis for $D > 0$. This comes from the analysis of the effective spin model derived from atomistic interactions in $\textrm{CrI}_3$, see Ref. \cite{Lado.Fernandes.2017}.

The convention with a minus sign standing at the beginning of \cref{eq:hamiltonian} was chosen to favor the ferromagnetic interactions for $J > 0$, and the alignment of spins in the off-plane axis for $D > 0$.

The main goal of this work is to analyze the groundstate of the above Hamiltonian by means of numerical calculations, for varying values of $J$, $\lambda$ and $D$.


\subsection{Classical approximation}

\begin{figure}[h]

    \begin{subfigure}{0.5\textwidth}
        \centering
        \includegraphics[width=0.9\linewidth]{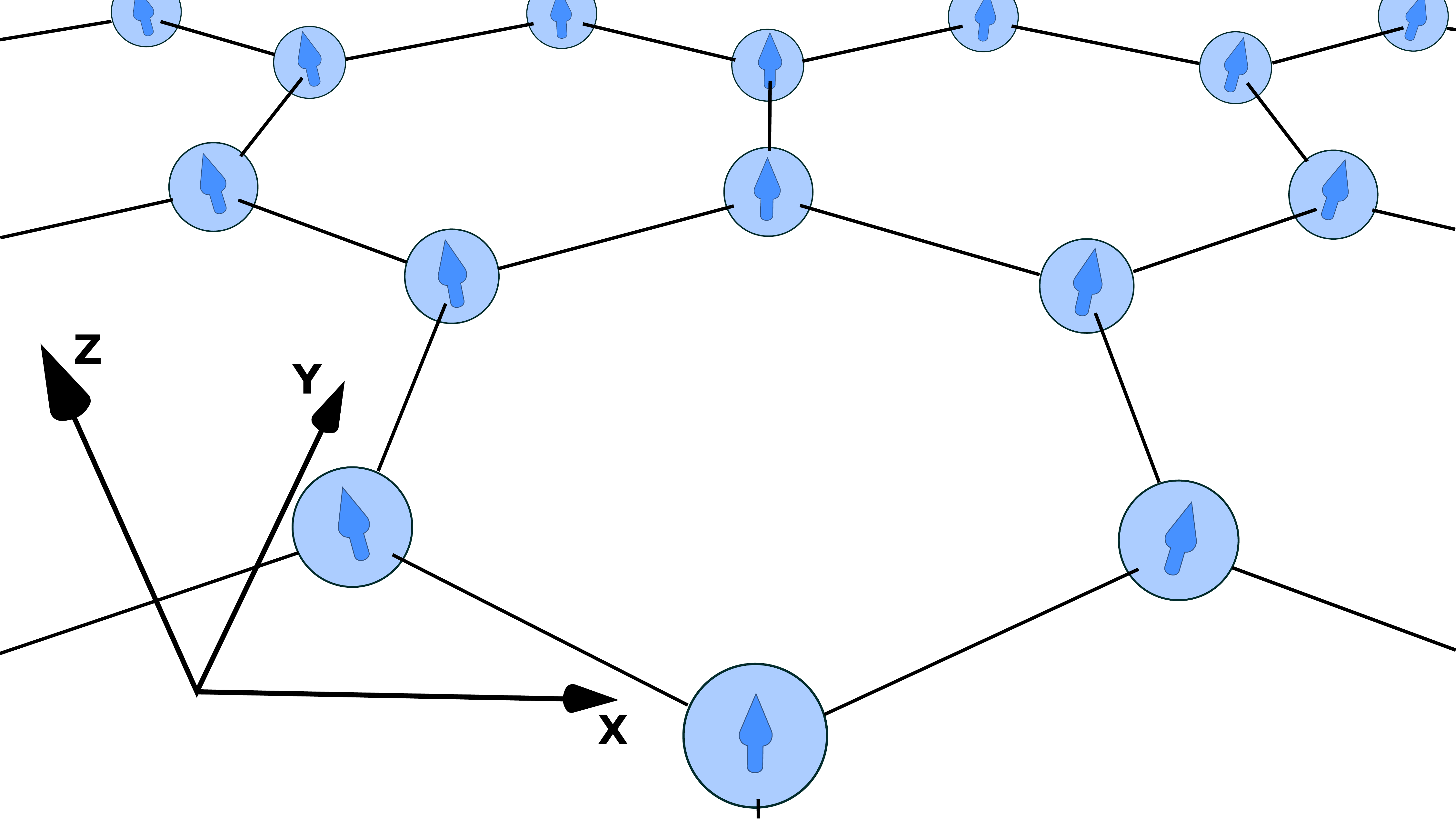} 
        \caption{}
        \label{fig:FM_Z}
    \end{subfigure}
    \begin{subfigure}{0.5\textwidth}
        \includegraphics[width=0.9\linewidth]{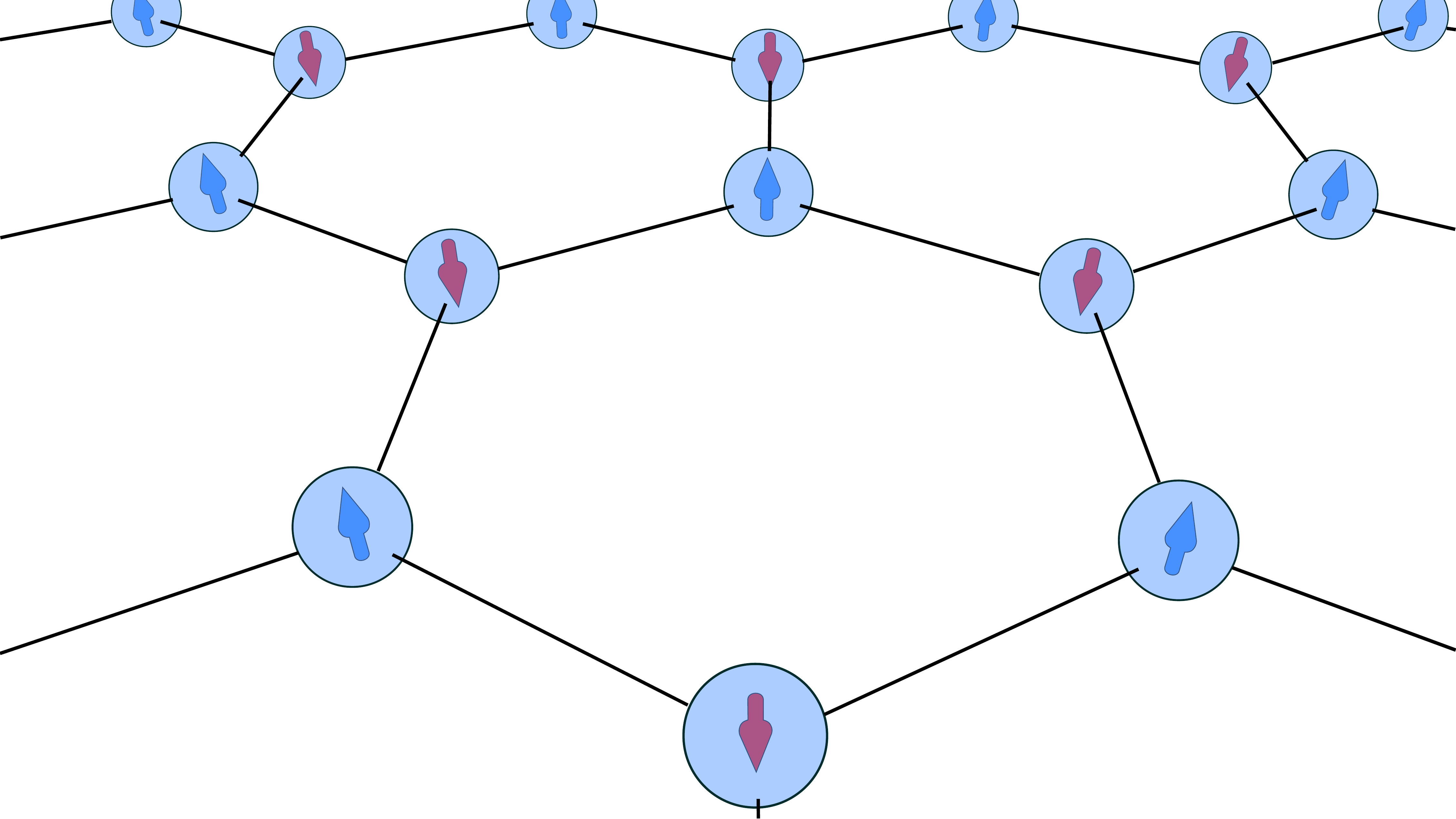}
        \caption{}
        \label{fig:AFM_Z}
    \end{subfigure}
    \newline
    \begin{subfigure}{0.5\textwidth}
        \centering
        \includegraphics[width=0.9\linewidth]{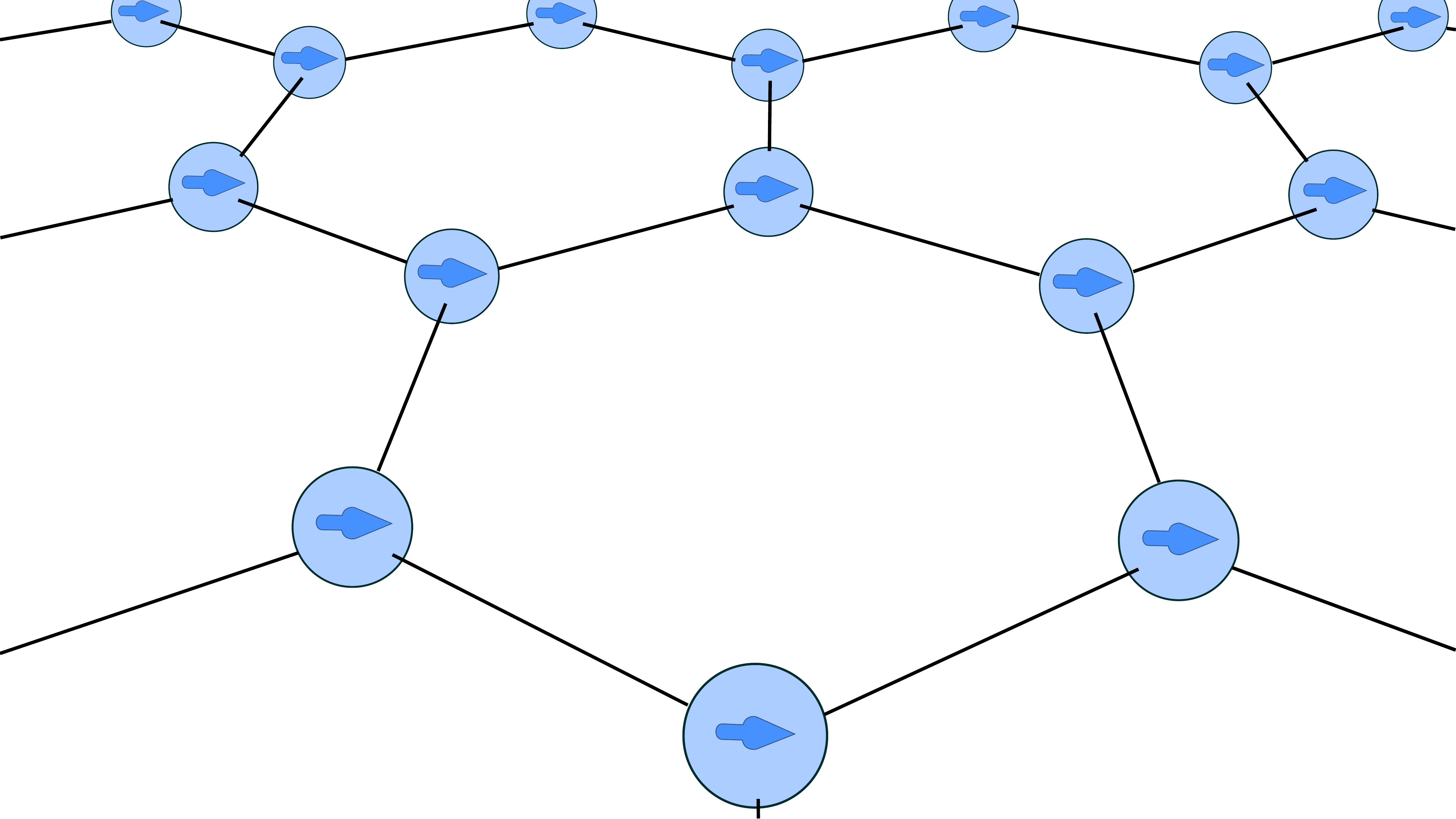} 
        \caption{}
        \label{fig:FM_XY}
    \end{subfigure}
    \begin{subfigure}{0.5\textwidth}
        \includegraphics[width=0.9\linewidth]{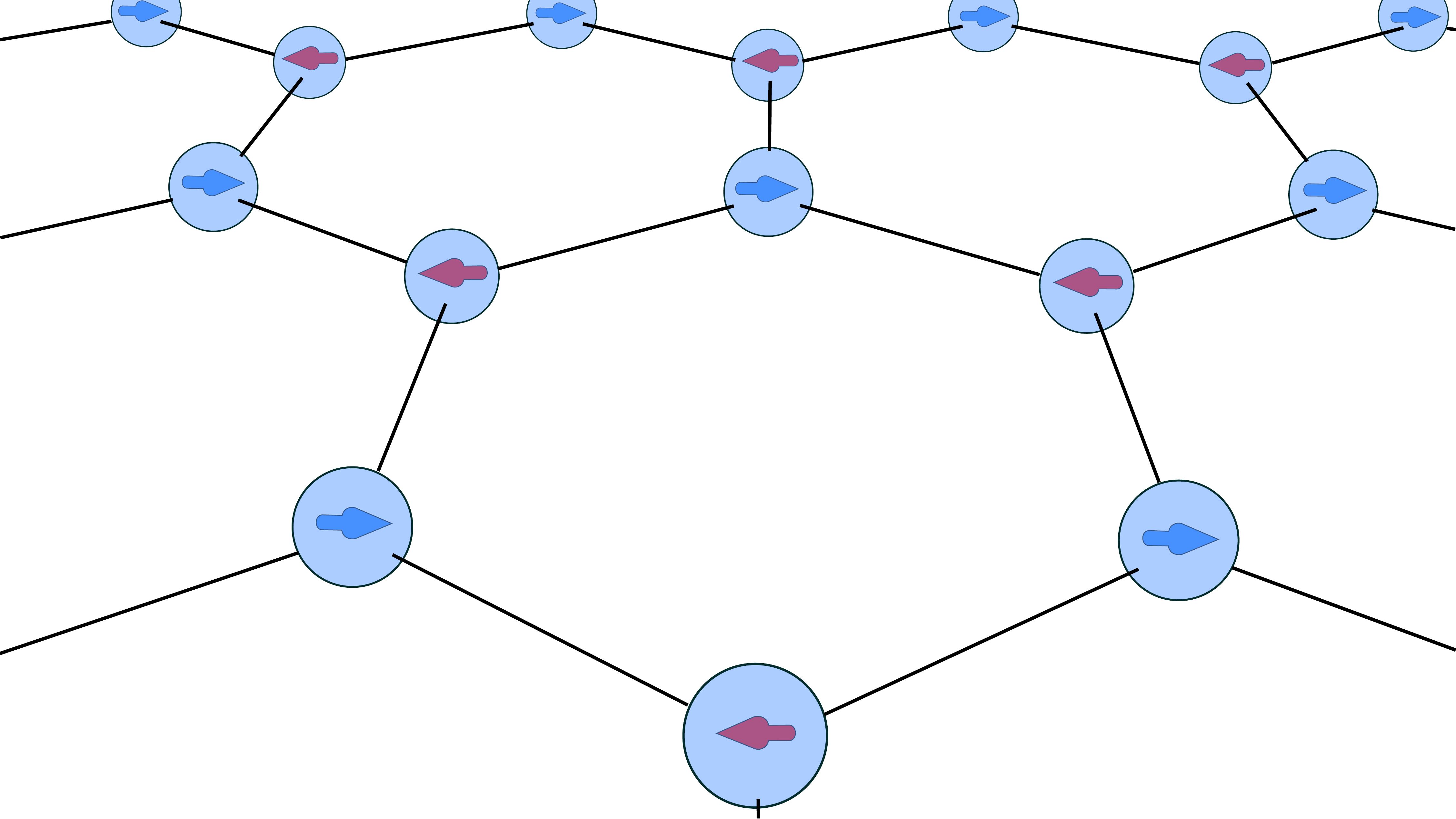}
        \caption{}
        \label{fig:AFM_XY}
    \end{subfigure}

    \caption{Four possible magnetic orders appearing in \cref{eq:hamiltonian}. (a) Ferromagnetic off-plane. (b) Antiferromagnetic off-plane. (c) Ferromagnetic in-plane. (d) Antiferromagnetic in-plane. Note, that figures (c) and (d) show only two out of infinitely many arrangements of spins in the XY plane. These two cases should be rather thought of as superpositions of all of the states, for which neighbouring spins (c) are aligned in the same direction, or (d) are aligned in the opposite directions.}
    \label{fig:orders}
\end{figure}

Hamiltonian proposed in \cref{eq:hamiltonian} can exhibit four types of magnetic order (depicted in \cref{fig:orders}): ferromagnetic off-plane, antiferromagnetic off-plane, ferromagnetic in-plane and antiferromagnetic in-plane. These phases compete with each other, when the system parameters are varied, with a ferromagnetic order favored for $J > 0$ and an antiferromagnetic order for $J < 0$.

To approximate the ground state energy, and at the same time the phase exhibited by the system, we prepared four classical product states. In the first one spins were aligned in parallel in the $Z$ axis, in the second one antiparallel in the $Z$ axis, in the third one parallel in the $X$ axis, and finally in the fourth case antiparallel in the $X$ axis. For each of these four states we calculated the expectation value of energy of the Hamiltonian given in \cref{eq:hamiltonian}, and picked the one with the lowest value as the ground state energy. 

\subsection{DMRG method} 

The DMRG method has been originally designed by Steven R. White \cite{White.1992,White.1993} for efficient numerical simulation of 1D quantum many-body gaped systems with short-range interactions. In the DMRG algorithm the Hamiltonian is projected onto the subspace of the whole Hilbert space, in which the low-lying eigenstates, in particular the ground state, are located. When increasing the size of the system, the size of the corresponding Hilbert space can be reduced by keeping only the significant values of the reduced density matrix of the entire system, when it is spatially divided into two smaller subsystems. DMRG is such a powerful method for these kinds of systems, because the groundstate entanglement entropy, associated with the eigenvalues of the reduced density matrix, increases according to the so-called area law \cite{Eisert.Cramer.2010}.

For a 1D system, edges of the system consist of just two ends of the chain, so its area is constant. This fact allows simulations to be carried out even for 1D chains of infinite length, because the projected Hilbert space has approximately constant dimension. DMRG has been further generalized to two and three dimensional systems, by the application of modern tensor network approaches, like Matrix Product States (MPS) \cite{Schollwock.2011}  and Projected Entangled-Pair States (PEPS) \cite{PEPS.2004}.
In such a case, the dimension of the projected Hilbert space is not constant, but still determined by the area law, which often allows to perform simulations of larger systems than the Exact Diagonalization (ED) method. In our calculations we were using the DMRG algorithm in the MPS formalism.

In order to investigate the properties of the XXZ model we are projecting the 2D honeycomb lattice onto a 1D spin chain, as depicted in \cref{fig:DMRG_lattice}. The system under study has periodic boundary conditions along both of the basis vectors. The Hamiltonian is divided into sectors with conserved quantum numbers. In our case, the conserved quantity is the total value of spin in the $Z$ axis. DMRG was executed separately on each of these sectors, to obtain the ground state energy and the energy of first two excited states. 

To check the scaling of the properties obtained from the finite DMRG in the torus geometry we repeated the calculations for a selected set of Hamiltonian's parameters using the infinite DMRG algorithm (iDMRG) in the infinite cylinder geometry.

\section{DMRG simulations at zero temperature} \label{sec:dmrg}

\begin{figure}
    \centering
    \includegraphics[scale=0.8]{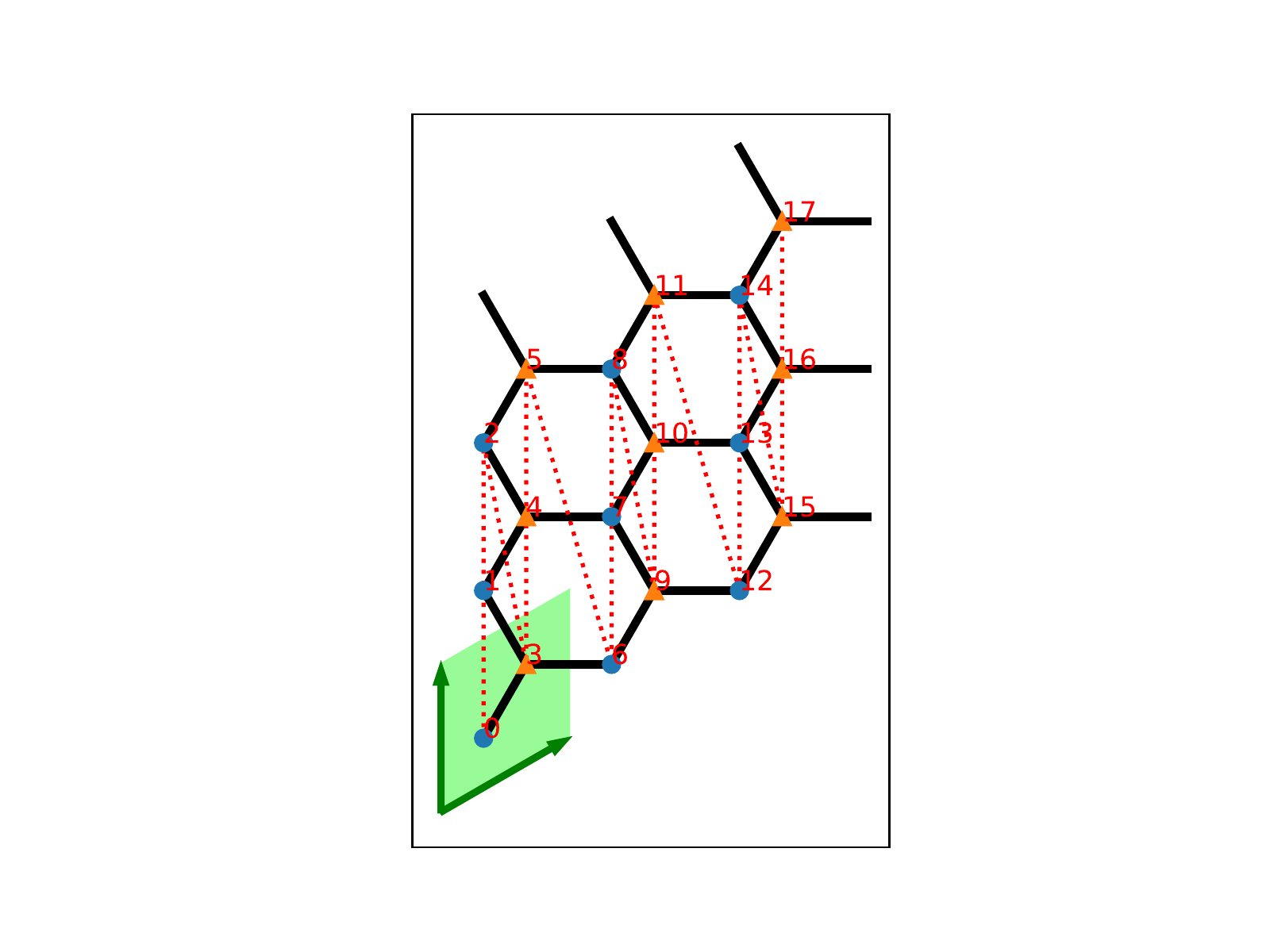}
    \caption{Honeycomb lattice of the system under study. Each node corresponds to a single Cr atom, which is represented by a spin $S = 3/2$ particle. The 2D system is projected onto a 1D spin chain in a way marked by a red, dotted line.}
    \label{fig:DMRG_lattice}
\end{figure}

We started our calculations with the case of isolated $\textrm{CrI}_3$, for which the parameters of the Hamiltonian were determined by means of DFT calculations \cite{Lado.Fernandes.2017}, and are equal to $D = 0$ meV, $J = 2.2$ meV and $\lambda = 0.09$ meV. As mentioned in the previous section, we focused on a honeycomb lattice in torus geometry depicted in \cref{fig:DMRG_lattice}, and chose 9 unit cells as the system's size, giving in total 18 spin $S=3/2$ sites. It should be noted that the full Hilbert space of such a system is equal to $4^{18}$, which is far beyond the capabilities of exact simulation methods. Because of that, we are allowing for an MPS's bond truncation during the execution of DMRG, with the maximally allowed error of $10^{-10}$. The ground state corresponds to a ferromagnetic order in the $Z$ axis with the total energy $E_{GS}=-139.1175$ meV, which is exactly equal to the classical energy. 
Subsequently, we modify Hamiltonian's parameters and investigate the ground state properties within the regimes of the four possible classical phases. Because the impact of a single ion anisotropy $D$ on the magnetic order in the whole system is significantly smaller than the one coming from the exchange interaction between particles \cite{Lado.Fernandes.2017} we focused mainly on the analysis of the properties of the system for varying values of parameters $J$ and $\lambda$, while fixing the value of $D$. We carried out calculations for $D \in \{ -0.4, 0.001, 0.4\}$ meV and $J, \lambda \in \{-0.5, -0.45, ..., 0.45, 0.5\}$ meV. In this analysis of the dependence of the magnetic phase on the selected set of parameters we restricted the lattice size to 4 unit cells while preserving the torus geometry, giving a total of 8 spin $S = 3/2$ particles.

\begin{figure}[h]
    \centering
    \includegraphics[width=\textwidth]{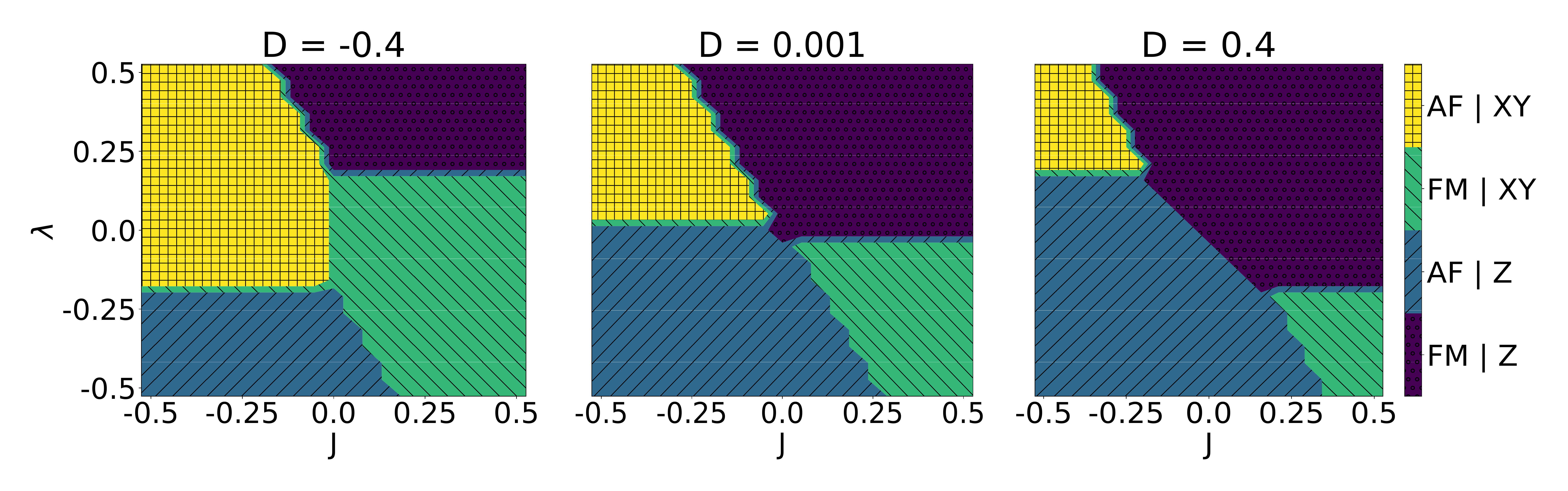}
    \caption{Magnetic phases predicted by the classical approximation.}
    \label{fig:classical_phase}
\end{figure}

\begin{figure}[h]
    \centering
    \includegraphics[width=\textwidth]{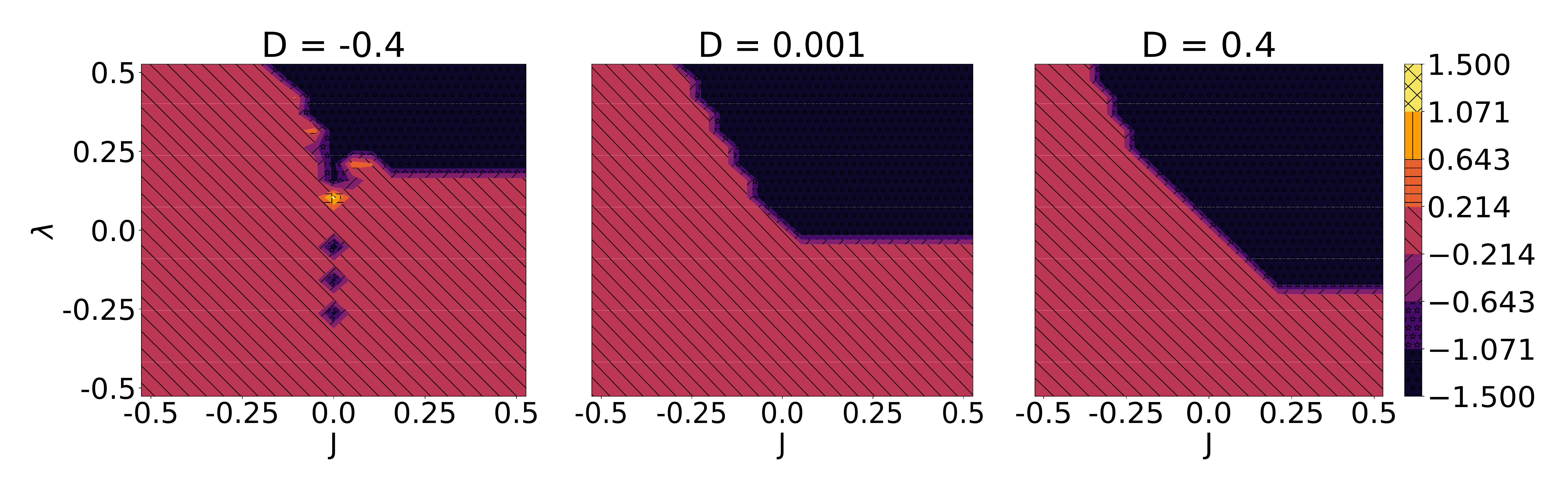}
    \caption{Average value of spin in the $Z$ axis.}
    \label{fig:average_Sz}
\end{figure}

\begin{figure}[h]
    \centering
    \includegraphics[width=\textwidth]{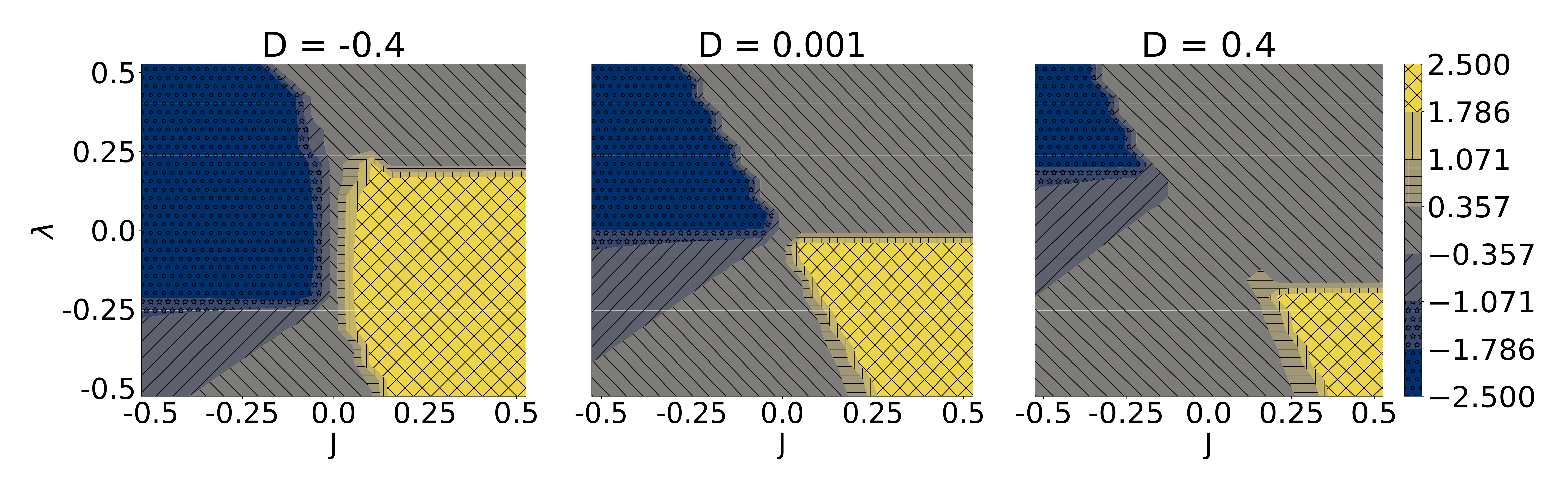}
    \caption{Average correlation in the $XY$ plane between the nearest neighbouring spins.}
    \label{fig:correlation}
\end{figure}

In \cref{fig:classical_phase} we present magnetic phases predicted by the classical approximation of the XXZ model, which are further compared with the results obtained from DMRG. The average value of spin in the $Z$ axis is shown in \cref{fig:average_Sz}, and in-plane correlations defined as $(\langle S_i^+ S_{i'}^- \rangle + \langle S_i^- S_{i'}^+\rangle )/2$, where indices $i$ and $i'$ correspond to the nearest neighbouring sites, are depicted in \cref{fig:correlation}. We note that it is not possible to measure the average value of spin in $X$ or $Y$ axes separately, as such measurements would always give $0$. Combinations of values from \cref{fig:average_Sz} and \cref{fig:correlation} give four possible outcomes. 

In the first case, the spins are fully aligned in the $Z$ axis in the absence of coexisting correlation in the $XY$ plane, resulting in ferromagnetic ordering in the off-plane axis (top-right corner of the parameter space). In \cref{fig:average_Sz} we see a black area, where the average spin value is equal to -1.5, which corresponds to a situation where all spins are fully aligned in the negative $Z$ axis direction. However, the ground state is degenerate, and a scenario where all spins are aligned in the positive off-plane axis direction is just as likely, and corresponds to the same energy value.

In the second case, both the average value of spin in the $Z$ axis and correlation in the $XY$ plane are equal to 0. After looking closely at the exact spin value at each site we noticed, that they are alternately arranged in opposite directions in the off-plane axis, thus staggered sublattice polarization is finite, which explains the average spin value equal to 0 in this direction and zero correlation between spins in the $XY$ plane. Therefore, these results correspond to an antiferromagnetic order in the $Z$ axis (bottom-left corner).

In the third case, the combination of the average value of spin in the $Z$ axis equal to 0 with positive correlation between neighbouring sites in the $XY$ plane (reaching a maximum value of 2.347) gives a ferromagnetic order in the $XY$ plane (bottom-right corner). 

The last, fourth case is analogous to the last one discussed, where the only difference is the negative correlation between spin sites (with a minimum value of -2.341), which corresponds to an antiferromagnetic order in the $XY$ plane (top-left corner). 

We confirm that the properties of the ground state of the fully quantum spin model are in agreement with classical predictions in the major part of the parameter space, for which calculations were conducted. Moreover, in line with the formalism adopted in the definition of the Hamiltonian \cref{eq:hamiltonian} it can be seen, that $D<0$ favors the in-plane phases, while $D>0$ favors the off-plane ones. Additionally, some noise at the phase transitions can be seen, which is related to convergence problems in DMRG and is not of physical origin.

\begin{figure}[h]
    \centering
    \includegraphics[width=\textwidth]{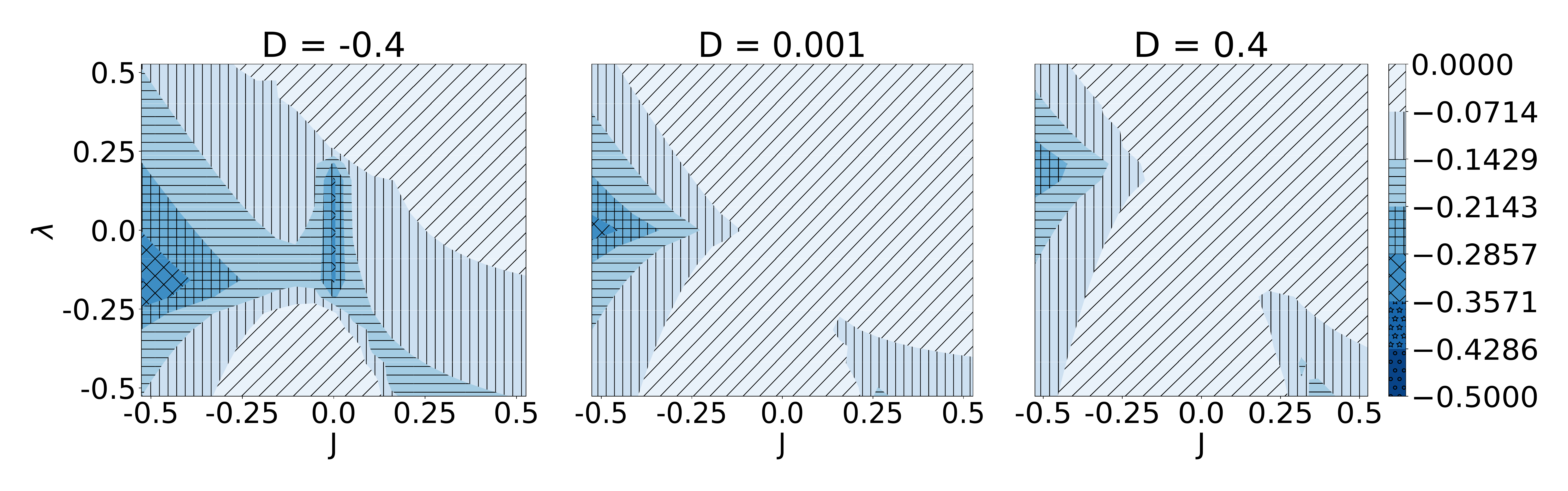}
    \caption{The correlation energy determined as a difference between the ground state obtained from DMRG and the classical approximation.}
    \label{fig:quantum_classical_differences}
\end{figure}

\textit{Energy differences between the quantum and classical models.} We can determine how well the classical approximation works by calculating the correlation energy $E_{corr}=E_{DMRG}-E_{clas}$, the differences between energies obtained from the DMRG calculations ($E_{DMRG}$) and classical approximations ($E_{clas}$). They are shown in \cref{fig:quantum_classical_differences}. The correlation energy is zero for the $Z$ axis ferromagnteic phase, and is the largest in magnitude for ferromagnetic and antiferromagnetic phases in the $XY$ plane. It increases gradually with the strength of $J$. Comparing  \cref{fig:quantum_classical_differences} with \cref{fig:average_Sz} and \cref{fig:correlation}, one can notice that the correlations are strongest close to phase boundaries, e.g. around $\lambda=0.2$ eV and $D=0.4$ eV. In \ref{app:appendix} the values of energy gap and average entanglement entropy are presented. The energy gaps are finite within the off-plane phases, while in-plane phases are gapless. Oppositely, the average entanglement entropy is finite within the in-plane phases and vanishes within the off-plane phases, with exception of the antiferromagnetic phase in the $Z$ axis, where a non-zero entanglement can be present. The maxima of the average entanglement entropy overlap with maxima of correlation energies.

\textit{Scaling of system's properties} For $D \in \{-0.4, 0.001, 0.4\}$ and picked values of $J$ and $\lambda \in \{-0.5, 0.5\}$ we repeated calculations using the iDMRG algorithm on an infinite cylinder. Aforementioned parameters correspond to all of the corners in the diagrams presented in this section. The circumference of the cylinder was equal to 2 sites, while one unit cell of the infinite MPS (iMPS) contained 4 sites. Of all off the subspaces with conserved total value of spin in the $Z$ axis, we chose only two - the one with $\langle S_z \rangle = 0$ and $\langle S_z \rangle = 3/2 \cdot N = 6$, where $N$ is the number of sites in the unit cell of the iMPS. We chose these subspaces, because we knew these were the regions where the groundstates were lying. Using (the same as in the case of the finite algorithm) maximally allowed truncation error of $10^{-10}$ and conducting at most 1000 sweeps of the iDMRG we found, that system's properties predicted by the finite algorithm are also reproduced by the infinite method. Results of the iDMRG for gradually increasing value of maximally allowed bond size $\chi$ are presented in \ref{app:appendix}.

\section{Monte Carlo simulations at finite temperature} \label{sec:monte_carlo}

In this section we would like to return to the case of the isolated $\textrm{CrI}_3$ and analyze its properties at finite temperatures. For that purpose we could also use tensor networks techniques, such as purification \cite{Zwolak.2004,Verstraete.2004,Feiguin.2005,Karrasch.2013,Bruognolo.2017} and the METTS (minimally entangled typical thermal state) algorithm \cite{Bruognolo.2017,White.2009,Stoudenmire.2010}. The purification method, by introducing the ancilla environment (which extends the considered Hilbert space), directly computes the thermal density matrix using imaginary time evolution. This approach works well for high temperatures, but its cost grows rapidly with the decrease of temperature. To diminish this issue METTS algorithm can be used, which blends imaginary
time evolution with Monte Carlo sampling. Therefore, to investigate the whole temperature range for the system under study we would have to combine purification with METTS, which in combination with large local Hilbert space of spin $S = 3/2$ particles would require huge computational overhead.

However, in the previous section we showed that classical approximation can be successfully used to describe the magnetic phase of the isolated $\textrm{CrI}_3$. This observation suggests, that also its thermodynamic properties could be efficiently studied by methods of classical physics, which prompted us to use Monte Carlo simulation \cite{Metropolis.1953, Landau.Binder.2009}, which is the most common method used to solve effective classical spin models of low-dimensional nanostructures \cite{Elyacoubi.Masrour.2018,Bahmad.Benyoussef.2007,Fadil.Mhirech.2019,Maaouni.Qajjour.2019,Zahraouy.Bahmad.2005,Masrour.Jabar.2018,Sahdane.Masrour.2021, Belhamra.Masrour.2020,Masrour.Jabar.2016,Jabar.Masrour.2018,Masrour.Jabar.2017,Jabar.Masrour.2017,Fadil.Maaouni.2020,Okabe.Kikuchi.1988, Cuccoli.Tognetti.1995,Jabar.Masrour1.2017}. In this work this technique was used to determine the Curie temperature and the type of phase transition. We note here, that the thermodynamics of a physical system depend on the full spectrum of its Hamiltonian, so the application of the classical approximation is not fully justified. Even though, results obtained in this section are in good agreement with experiments.


\begin{figure}[h]
    \centering
    \includegraphics[scale=0.55]{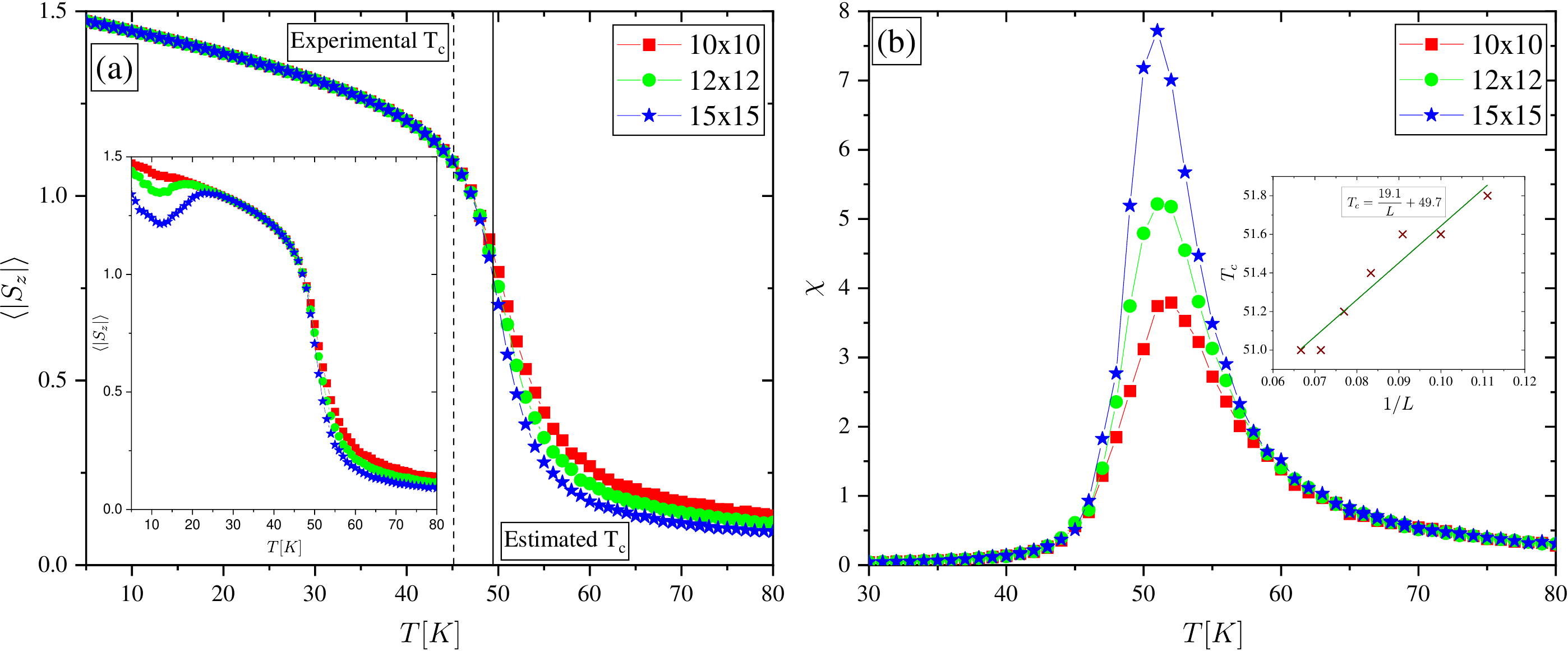}
    \caption{(a) Relationship between $\langle |S_z| \rangle$ and $T$ for ordered initial conditions. Inset: The same dependency, but for random initial conditions. (b) Magnetic susceptibility as a function of temperature. Inset: Scaling of the critical temperature.}
    \label{fig:MC_CrI3}
\end{figure}


The average spin $\langle|S_z|\rangle$ value as a function of temperature obtained from MC calculations, for the parameter set predicted for the isolated $\textrm{CrI}_3$, is shown in \cref{fig:MC_CrI3} (a). We focused on three different systems consisting of $L \times L$ unit cells (one unit cell contained two spins), with periodic boundary conditions along both of the basis vectors. The calculations were conducted for $L=10, 12$ and 15, yielding 200, 288 and 450 spin particles, respectively.
The simulation was performed on two types of initial conditions (IC): random and ordered ones. 
For the ordered IC, each spin was oriented straight along the $Z$-axis ($S_z = 3/2$), while for the random IC the orientation of spins was initiated by uniformly distributed random numbers.
Results for the ordered IC are shown in the main plot of \cref{fig:MC_CrI3} (a), while the ones for random IC are depicted in the inset.
The discontinuous phase transition is characterized by the existence of the hysteresis, in contrast to the continuous one.
There are no visible differences between the two types of IC, which strongly suggests, that the hysteresis in the system under consideration does not exist, so the phase transition is continuous.

The Curie temperature $T_c$ in the MC simulation can be easily determined by finding the maximum value of magnetic susceptibility $\chi$, which in turn can be obtained as a magnetization standard deviation \cite{Landau.Binder.2009}
\begin{equation}
    \chi = \frac{N}{k_B T} \left( \left< S_z^2\right> - \left< S_z\right>^2  \right),
\end{equation}
where $N$ is the total number of spins in the system (here $N = 2L^2$).
The relationship between magnetic susceptibility $\chi$ and temperature is shown in \cref{fig:MC_CrI3} (b). 
Peaks of $\chi$ are visible near the experimental value of $T_c = 45 K$.
The inset in \cref{fig:MC_CrI3} (b)  shows the scaling of the position of $\chi$ peaks with the inverse length $1/L$, and has been interpolated by the linear regression. 
In the thermodynamic limit $1/L \xrightarrow{} 0$, the peak position corresponds to the Curie temperature $T_c$ of the macroscopic system, and it has been estimated at $T_c ~= 49.7 K$.
This result is closer to the value obtained in the experiment than previously recorded lower-bound of $33K$, and upper-bound of $85K$ \cite{Lado.Fernandes.2017}.

\section{Conclusions} \label{sec:conclusions}

$\textrm{CrI}_3$ can exhibit various types of magnetic ordering by introducing defects \cite{Pizzochero.2020}, strain \cite{Webster.Yan.2018,Zheng.Zhao.2018} and charge doping \cite{Zheng.Zhao.2018,Jiang.Li.2018}. We analyzed the role of correlation effects and studied phase transitions in the XXZ model on a honeycomb lattice, which is an effective model for $\textrm{CrI}_3$. For this purpose we used the finite DMRG method on a lattice with torus geometry, and iDMRG on an infinite cylinder. We found, that magnetic order realized in the ground state of the XXZ model can be predicted by classical means with high precision. The correlation energy is zero for the off-plane ferromagnetic phase, and is the largest in  magnitude for the in-plane ferromagnetic and antiferromagnetic phases. 

Our results make it possible to find such a set of $J$, $\lambda$ and $D$ parameters for which the energy gap is the largest, making the resulting phase more stable. The combination of large, positive $D$ with $J$ and $\lambda$, being both positive or negative, give the highest stability. Moreover, by combining results from \cref{fig:quantum_classical_differences} and \cref{fig:entanglement}, it is possible to choose such a set of Hamiltonian's parameters, where large values of entanglement entropy overlap with maximal correlation energies. Thanks to this, it is possible to obtain a strongly correlated, non-classical quantum state, which is of great interest.

We performed classical MC calculations on a lattice with torus geometry to determine the Curie temperature for $\textrm{CrI}_3$, and obtained the value $T_c = 49.7K$, which is in good agreement with experiment ($45K$).

\ack{
This work was supported by the National Science Centre (NCN, Poland) under grants:
2019/33/N/ST3/03137 (M.K.). 
Additionally, B.R. acknowledges support by the European Union under the European Social Fund. 
All calculations were performed using the TeNPy Library (version 0.7.2) \cite{tenpy} at the Wrocław Center for Networking and Supercomputing.
}
 
\section*{Conflict of Interest Statement}

All authors declare that they have no conflicts of interest.
 
\section*{Data availability statement}

All data that support the findings of this study are included within the article (and any supplementary files).
 
\appendix \section{} \label{app:appendix}

\begin{figure}[h]
    \centering
    \includegraphics[width=\textwidth]{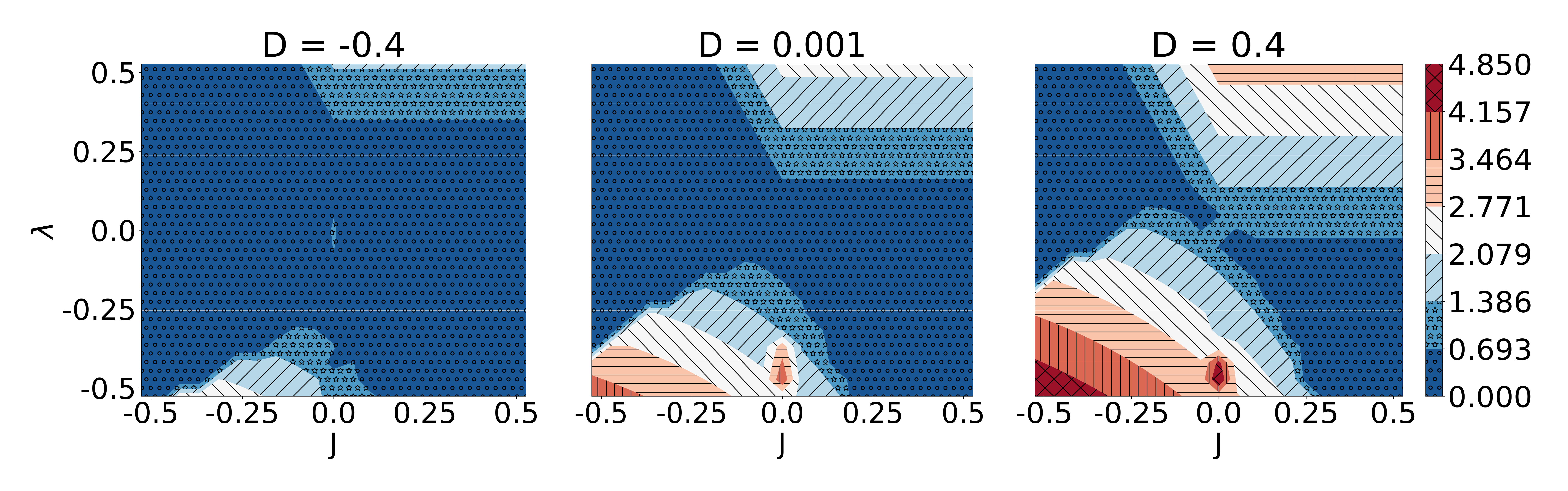}
    \caption{Energy gap between the ground state and first excited state.}
    \label{fig:energy_gap}
\end{figure}

\textit{Energy gap.} The differences in energy between the ground state and the first excited one are depicted in \cref{fig:energy_gap}. Increasing the value of $D$ opens the energy gap in the system (especially for large values of $J$ and $\lambda$, where both are positive or negative at the same time), giving the maximum value of 4.85 meV for $J = \lambda = -0.5$ meV and $D = 0.4$ meV.

\begin{figure}[h]
    \centering
    \includegraphics[width=\textwidth]{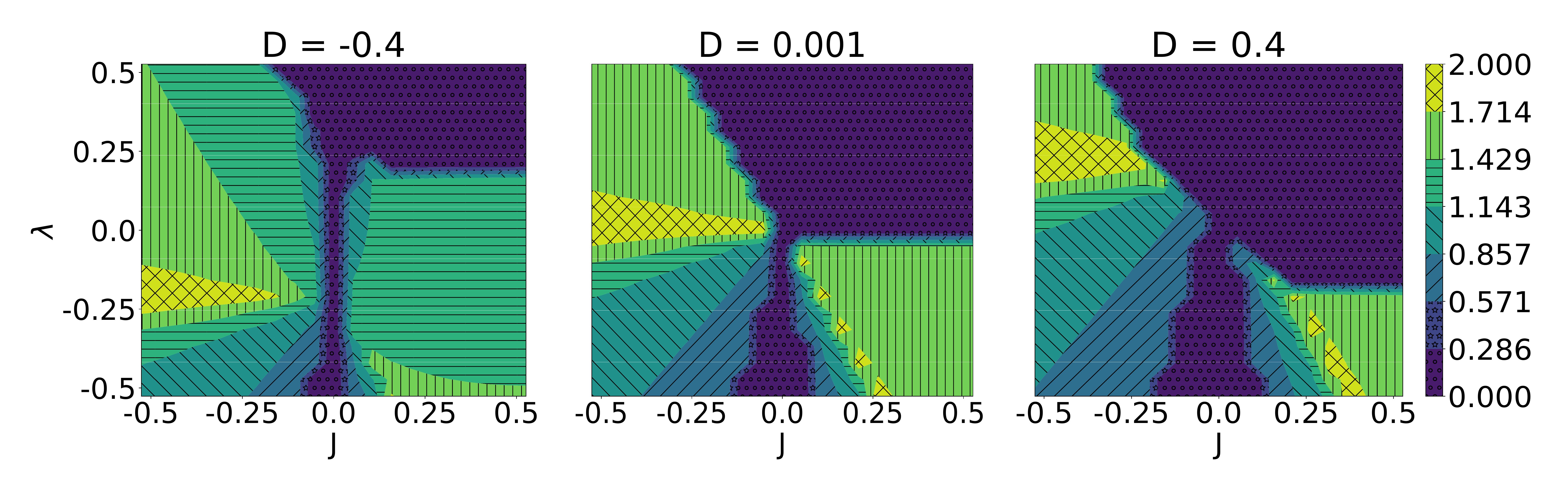}
    \caption{Average entanglement entropy in the XXZ model.}
    \label{fig:entanglement}
\end{figure}

\textit{Entanglement entropy.} We measured the (half-chain) von Neumann entanglement entropy for all nontrivial bonds of the ground state MPSs. Subsequently, we averaged these values over all conducted partitions of the lattice. Fig. \ref{fig:entanglement} depicts the average entanglement entropy observed in the system under study. It can be seen that, as expected, its largest values are obtained for the in-plane phases, reaching the maximum value of 1.903. However, we can also see some entanglement generated in the lattice in the antiferromagnetic phase in the $Z$ axis, which gradually increases as we approach the phase boundaries.

\textit{Scaling of system's properties} In \cref{tab:mean_energy,tab:mean_spin_Z,tab:mean_correlations_XY} we compare results obtained from DMRG and iDMRG methods. We launched iDMRG multiple times with increasing value of maximally allowed bond size $\chi$. We can see, that the properties of the system's groundstate are consistent between the two methods used and also do not change with the increasing value of $\chi$. It should be noted, that varying values of $\chi$ gave different excited states, but in this work we are interested only in the groundstate of the system.

\newpage
\begin{table}[!ht]
    \centering
    \begin{tabular}{|c|c|c|c|c|c|c|}
    \hline
        \multirow{2}{3em}{D} & \multirow{2}{3em}{J} & \multirow{2}{3em}{$\lambda$} & \multirow{2}{3em}{DMRG} & \multicolumn{3}{|c|}{iDMRG} \\ \cline{5-7}
        & & & & $\chi = 1000$ & $\chi = 2000$ & $\chi = 3000$ \\ \hline
        \multirow{4}{3em}{-0.4} & -0.5 & -0.5 & -2.61813 & -2.61394 & -2.61394 & -2.61394 \\
        & -0.5 & 0.5 & -1.52755 & -1.49697 & -1.49697 & -1.49697 \\
        & 0.5 & -0.5 & -1.52755 & -1.49697 & -1.49697 & -1.49697 \\
        & 0.5 & 0.5 & -2.475 & -2.475 & -2.475 & -2.475 \\ \hline
        \multirow{4}{3em}{0.001} & -0.5 & -0.5 & -3.48795 & -3.48618 & -3.48618 & -3.48618 \\
        & -0.5 & 0.5 & -1.78903 & -1.76248 & -1.76248 & -1.76248 \\
        & 0.5 & -0.5 & -1.78903 & -1.76248 & -1.76248 & -1.76248 \\
        & 0.5 & 0.5 & -3.37725 & -3.37725 & -3.37725 & -3.37725 \\ \hline
        \multirow{4}{3em}{0.4} & -0.5 & -0.5 & -4.36585 & -4.36492 & -4.36492 & -4.36492 \\
        & -0.5 & 0.5 & -2.10218 & -2.07989 & -2.07989 & -2.07989 \\
        & 0.5 & -0.5 & -2.10218 & -2.07989 & -2.07989 & -2.07989 \\
        & 0.5 & 0.5 & -4.275 & -4.27499 & -4.27499 & -4.27499 \\ \hline
    \end{tabular}
    \caption{Mean energy per node.}
    \label{tab:mean_energy}
\end{table}

\begin{table}[!ht]
    \centering
    \begin{tabular}{|c|c|c|c|c|c|c|}
    \hline
        \multirow{2}{3em}{D} & \multirow{2}{3em}{J} & \multirow{2}{3em}{$\lambda$} & \multirow{2}{3em}{DMRG} & \multicolumn{3}{|c|}{iDMRG} \\ \cline{5-7}
        & & & & $\chi = 1000$ & $\chi = 2000$ & $\chi = 3000$ \\ \hline
        \multirow{4}{3em}{-0.4} & -0.5 & -0.5 & 0 & 0 & 0 & 0 \\
        & -0.5 & 0.5 & 0 & 0 & 0 & 0 \\
        & 0.5 & -0.5 & 0 & 0 & 0 & 0 \\
        & 0.5 & 0.5 & 1.5 & 1.5 & 1.5 & 1.5 \\ \hline
        \multirow{4}{3em}{0.001} & -0.5 & -0.5 & 0 & 0 & 0 & 0 \\
        & -0.5 & 0.5 & 0 & 0 & 0 & 0 \\
        & 0.5 & -0.5 & 0 & 0 & 0 & 0 \\
        & 0.5 & 0.5 & 1.5 & 1.5 & 1.5 & 1.5 \\ \hline
        \multirow{4}{3em}{0.4} & -0.5 & -0.5 & 0 & 0 & 0 & 0 \\
        & -0.5 & 0.5 & 0 & 0 & 0 & 0 \\
        & 0.5 & -0.5 & 0 & 0 & 0 & 0 \\
        & 0.5 & 0.5 & 1.5 & 1.5 & 1.5 & 1.5 \\ \hline
    \end{tabular}
    \caption{Average value of spin in the $Z$ axis.}
    \label{tab:mean_spin_Z}
\end{table}

\newpage
\begin{table}[!ht]
    \centering
    \begin{tabular}{|c|c|c|c|c|c|c|}
    \hline
        \multirow{2}{3em}{D} & \multirow{2}{3em}{J} & \multirow{2}{3em}{$\lambda$} & \multirow{2}{3em}{DMRG} & \multicolumn{3}{|c|}{iDMRG} \\ \cline{5-7}
        & & & & $\chi = 1000$ & $\chi = 2000$ & $\chi = 3000$ \\ \hline
        \multirow{4}{3em}{-0.4} & -0.5 & -0.5 & -0.42120 & -0.39787 & -0.39787 & -0.39787 \\
        & -0.5 & 0.5 & -2.35530 & -2.33065 & -2.33065 & -2.33065 \\
        & 0.5 & -0.5 & 2.35530 & 2.33065 & 2.33065 & 2.33065 \\
        & 0.5 & 0.5 & 0 & 0 & 0 & 0 \\ \hline
        \multirow{4}{3em}{0.001} & -0.5 & -0.5 & -0.31161 & -0.30305 & -0.30305 & -0.30305 \\
        & -0.5 & 0.5 & -2.38443 & -2.35844 & -2.35844 & -2.35844 \\
        & 0.5 & -0.5 & 2.38443 & 2.35844 & 2.35844 & 2.35844 \\
        & 0.5 & 0.5 & 0 & 0 & 0 & 0 \\ \hline
        \multirow{4}{3em}{0.4} & -0.5 & -0.5 & -0.25103 & -0.24679 & -0.24679 & -0.24679 \\
        & -0.5 & 0.5 & -2.34179 & -2.31411 & -2.31411 & -2.31411 \\
        & 0.5 & -0.5 & 2.34179 & 2.31411 & 2.31411 & 2.31411 \\
        & 0.5 & 0.5 & 0 & 0 & 0 & 0 \\ \hline
    \end{tabular}
    \caption{Average correlation in the $XY$ plane.}
    \label{tab:mean_correlations_XY}
\end{table}

\section*{References}
\bibliography{bib.bib}

\end{document}